# Seebeck-driven transverse thermoelectric generation in magnetic hybrid bulk materials

Weinan Zhou[1], Asuka Miura[1,a)], Takamasa Hirai[1], Yuya Sakuraba[1], and Ken-ichi Uchida[1,2,b)]

**AFFILIATIONS**

[1] National Institute for Materials Science, Tsukuba 305-0047, Japan

[2] Institute for Materials Research, Tohoku University, Sendai 980-8577, Japan

a) **Present address:** Integrated Research for Energy and Environment Advanced Technology, Kyushu Institute of Technology, Kitakyushu, Fukuoka 804-8550, Japan

b) **Authors to whom correspondence should be addressed:** UCHIDA.Kenichi@nims.go.jp

**ABSTRACT**

The Seebeck-driven transverse thermoelectric generation in magnetic/thermoelectric hybrid materials (STTG) has been investigated in all-bulk hybrid materials. The transverse thermopower in a ferromagnetic $Co_2MnGa$/thermoelectric $n$-type Si hybrid bulk material with the adjusted dimensions reaches 16.0 μV/K at room temperature with the aid of the STTG contribution, which is much larger than the anomalous Nernst coefficient of the $Co_2MnGa$ slab (6.8 μV/K). Although this transverse thermopower is smaller than the value for previously reported thin-film-based hybrid materials, the hybrid bulk materials exhibit much larger electrical power owing to their small internal resistance. This demonstration confirms the validity of STTG in bulk materials and clarifies its potential as a thermal energy harvester.

The thermoelectric generation is one of the promising technologies to realize a sustainable society because it directly converts heat into electricity in a solid. The conventional thermoelectric generation is driven by the Seebeck effect, in which an electric field **E** is generated in the direction parallel to a temperature gradient $\nabla T$. Because of the parallel relationship between **E** and $\nabla T$, the Seebeck effect is classified as a longitudinal thermoelectric effect. Due to the longitudinal geometry, a thermoelectric module based on the Seebeck effect typically consists of many pairs of $p$-type and $n$-type semiconductors alternately arranged and connected in series. Such a complicated three-dimensional structure comprising many junctions often causes problems, e.g., low mechanical durability, high production cost, efficiency losses in contact resistances, and deterioration of the hot side under a large temperature difference.[1] These issues hinder wide range applications of thermoelectric technologies.

Thermoelectric technologies are now reaching a turning point, and the key to this transition is transverse thermoelectric effects, in which **E** is generated in the direction perpendicular to $\nabla T$.[2-16] Owing to the orthogonal relation between **E** and $\nabla T$, the output voltage (output electrical power) induced by the transverse thermoelectric effects can be enhanced simply by elongating the length (enlarging the area) of a material perpendicular to applied $\nabla T$ without forming a complicated three-dimensional structure. The transverse thermoelectric effects are thus suitable for harvesting thermal energy distributed over a large area. Significantly, the junction-less structure promises to overcome the above problems that limit applications of the Seebeck module, potentially bringing the efficiency of thermoelectric modules close to the theoretical value.[14-16] To utilize such advantages, many studies have focused on developing physics and applications of transverse thermoelectric phenomena that appear due to magnetic fields/magnetization[3,7-13] and to intrinsic/artificial anisotropic transport properties.[2,4-6,14,15] Particularly, with the recent development of topological materials science[17] and spin caloritronics,[18-20] research on the anomalous Nernst effect (ANE) is rapidly progressing,[9-13] where **E** is generated in the cross-product direction of $\nabla T$ and the spontaneous magnetization **M** in a magnetic material [Fig. 1(a)].

As an effort to improve the transverse thermoelectric conversion performance using magnetic materials, another mechanism called the Seebeck-driven transverse thermoelectric generation in magnetic/thermoelectric hybrid materials (STTG) has been demonstrated in 2021.[21] Although ANE appears in a plain magnetic material, STTG requires a hybrid material comprising a magnetic metal and a thermoelectric semiconductor,[21-23] where the magnetic metal and thermoelectric semiconductor show the large anomalous Hall effect (AHE)[24] and large Seebeck effect, respectively. Figure 1(b) shows a schematic of the typical hybrid material used for STTG. When $\nabla T$ is applied to a closed circuit comprising the magnetic and thermoelectric materials, a charge current is induced by the difference in the Seebeck coefficients of the materials. This charge current is in turn converted into a transverse electric field owing to AHE in the magnetic material. The total transverse thermopower, i.e., transverse electric field normalized by longitudinal $\nabla T$, due to ANE and STTG is phenomenologically expressed as[21,22]

$$S_{\text{tot}}^{y} = S_{\text{ANE}} - \frac{\rho_{\text{AHE}}}{\rho_{\text{TE}}/r + \rho_{\text{M}}}(S_{\text{TE}} - S_{\text{M}}), \qquad (1)$$

where $S_{\text{ANE}}$ is the anomalous Nernst coefficient of the magnetic material, $\rho_{\text{AHE}}$ the anomalous Hall resistivity of the magnetic material, $\rho_{\text{M(TE)}}$ and $S_{\text{M(TE)}}$ the longitudinal resistivity and Seebeck coefficient of the magnetic (thermoelectric) material, respectively, and $r = (L_{\text{M}}^{x}/L_{\text{TE}}^{x}) \times (L_{\text{TE}}^{y}L_{\text{TE}}^{z}/L_{\text{M}}^{y}L_{\text{M}}^{z})$ the size ratio determined by the length of the magnetic (thermoelectric) material $L_{\text{M(TE)}}^{x,y,z}$ along the $x$, $y$, and $z$ directions. Here, we assume the situation that the electrodes for measuring the transverse thermopower are attached to the magnetic material in the $y$ direction and $\nabla T$ is applied along the $x$ direction. The second term on the right-hand side of Eq. (1) represents the STTG contribution and can reach the order of 100 µV/K by optimizing the combination of the magnetic and thermoelectric materials as well as the size ratio. In fact, the previous experiments show that the Co$_2$MnGa (CMG)/$n$-type Si hybrid material, in which CMG has large $\rho_{\text{AHE}}$ and Si has large $S_{\text{TE}}$, exhibits a transverse thermopower of 82.3 µV/K at room temperature.[21] This value is an order of magnitude larger than the record-high $S_{\text{ANE}}$ value at room temperature and quantitatively consistent with the prediction based on Eq. (1), indicating the usefulness and potential of STTG. However, at present, such a large thermopower is obtained only when the combination of magnetic films and thermoelectric slabs is used, where $r > 10000$. The hybrid materials based on magnetic thin films are not suitable for thermoelectric power generation because of the large internal resistance of the films. Thus, it is necessary to realize large STTG in all-bulk hybrid materials with reasonable $r$ values.

In this study, we show that STTG occurs even in all-bulk hybrid materials, where the magnetic material is also a slab and the $r$ value is much smaller than that for the previously reported thin-film-based materials. According to Eq. (1), the magnitude of the transverse thermopower due to STTG is monotonically decreased by decreasing $r$, i.e., increasing the thickness of the magnetic material relative to that of the thermoelectric material, when the other parameters are fixed. Nevertheless, we found that the STTG-induced thermopower can be much larger than the ANE-induced thermopower even in magnetic hybrid bulk materials. We demonstrated this by using CMG/$n$-type Si hybrid bulk materials and observed the maximum transverse thermopower of 16.0 µV/K at room temperature due to the STTG contribution, which is several times larger than $S_{\text{ANE}}$ of our CMG slab (6.8 µV/K). Although this thermopower is smaller than the value obtained in the thin-film-based materials, our magnetic hybrid bulk materials exhibit much larger electrical power owing to their small internal resistance. This result extends the scope of STTG to bulk materials and opens the possibility of thermal energy harvesting applications of the transverse thermoelectric conversion using magnetic materials.

First, we show that all-bulk hybrid materials are suitable for extracting power through STTG based on the phenomenological calculation. In accordance with the condition and formulation in Refs. 21 and 22, the effective electrical resistivity and resistance along the $y$ direction in the closed circuit comprising the magnetic and thermoelectric materials are respectively expressed as

$$\rho_{\text{eff}}^y = \rho_{\text{M}} + \frac{\rho_{\text{AHE}}^2}{\rho_{\text{TE}}/r + \rho_{\text{M}}}, \quad (2)$$

$$R_{\text{eff}}^y = \rho_{\text{eff}}^y \left( L_{\text{M}}^y / L_{\text{M}}^z L_{\text{M}}^x \right). \quad (3)$$

By using Eq. (1), the maximum output power is calculated to be

$$P_{\max} = L_M^x L_M^y L_M^z \frac{(S_{\text{tot}}^y \nabla T)^2}{4\rho_{\text{eff}}^y}. \tag{4}$$

This is obtained when the load resistance $R_L$ is optimized to be matched with $R_{\text{eff}}^y$ [see the equivalent circuit in Fig. 2(b)]. Figures 2(a) and (b) respectively show the $r$ dependence of $\rho_{\text{eff}}^y$ and $P_{\max}$ for the CMG/Si hybrid materials, calculated using the parameters for CMG and $n$-type Si used in this study (Table 1). We found that $\rho_{\text{eff}}^y$ is mainly determined by $\rho_M$ and weakly dependent on $r$. Thus, $P_{\max}$ is almost proportional to the volume of the magnetic material, and the output power monotonically increases with decreasing $r$, indicating that all-bulk hybrid materials with small $r$ values can generate larger power through STTG than thin-film-based systems.

The schematic and photograph of the CMG/Si hybrid bulk material used in this study are shown in Figs. 3(a) and (b), respectively. To exploit STTG in magnetic hybrid bulk materials with small $r$, the difference in the electrical resistivities between the magnetic and thermoelectric materials should be as small as possible, while $\rho_{\text{TE}}$ is four orders of magnitude larger than $\rho_M$ in the previously used thin-film-based samples.[21] We thus used a highly antimony-doped $n$-type Si substrate as a thermoelectric material, which is commercially available from Crystal Base Co., Ltd. The $\rho_{\text{TE}}$ value of the antimony-doped Si substrate is two orders of magnitude smaller than that of the phosphorus-doped Si used in Ref. 21, while $S_{\text{TE}}$ of the substrate used here is about half of that of the phosphorus-doped Si (Table 1). The dimensions of the Si substrate are $L_{\text{TE}}^x = 10$ mm, $L_{\text{TE}}^y = 6$ mm, and $L_{\text{TE}}^z = 1$ mm and its top surface is thermally oxidized. To reduce contact resistance, metal electrodes were attached to the side surfaces in the $y$-$z$ plane of the Si substrate using ultrasonic solder [Fig. 3(a)]. The CMG slab used as a magnetic material was prepared by the following processes. First, 99.97% purity Co, 99.99% purity Mn, and 99.9999% purity Ga shots, available from RARE METALLIC Co., Ltd., were balanced with an atomic ratio of 2:1:1 and arc-melted in an Ar atmosphere. The arc-melted ingot was annealed in a high vacuum at 1000 °C for 48 h and subsequently at 600 °C for 72 h for homogenization. The CMG ingot was crushed using a mortar and a planetary ball mill, followed by sieved through a 100 μm mesh. Next, a cylindrical CMG ingot of the 20 mm diameter and ~12 mm thickness was produced by sintering the prepared CMG powder at 800 °C and 30 MPa for 15 minutes under vacuum by a spark plasma sintering method. The sintered ingot was cut into rectangular-shaped slabs of a length of 9 mm and width of $L_M^y = 4$ mm. We prepared 5 CMG slabs with different thicknesses ranging from $L_M^z = 0.06$ mm to 0.61 mm using a mechanical polishing system, where all the slabs were cut from the same ingot. The composition of the slab was evaluated using an inductively coupled plasma optical emission spectrometer and determined to be $Co_{52.2}Mn_{22.6}Ga_{25.2}$, which is slightly Co-rich in comparison with the stoichiometric composition. The CMG slab is polycrystalline and its grain crystal orientations are random, showing the isotropic properties.

Nevertheless, our CMG slab exhibits large $S_{\text{ANE}}$ and $\rho_{\text{AHE}}$ comparable to the values for single-crystalline Co$_2$MnGa[9,12,21] (Table 1). The CMG slabs were then fixed on the thermally oxidized surface of the Si substrates with a thermally conductive silicone adhesive (COM-G52, COM Institute Inc.) and electrically connected to the electrodes on the sides of Si through pure indium. Since indium covers a part of the top surface of CMG near both the ends and the electrical conductivity of indium is much larger than that of CMG, we determined $L_{\text{M}}^{x}$ as the distance between the two indium contacts for each sample [Fig. 3(a)]. The resultant $r$ value in the CMG/Si hybrid bulk materials ranges from 18.0 to 1.9, more than three orders of magnitude smaller than that for the previously reported thin-film-based hybrid materials. The transverse thermopower along the $y$ direction between the edges of the CMG slab in the CMG/Si hybrid bulk materials was measured at room temperature with applying a uniform $\nabla T$ along the $x$ direction and sweeping an external magnetic field $H$ along the $z$ direction, where the thermopower probes were attached around the center of CMG [Fig. 3(a)]. We confirmed the uniformity of $\nabla T$ by monitoring the temperature distribution on the surface of the CMG slab using an infrared camera and estimated the magnitude of $\nabla T$ by dividing the temperature difference between the two points by their distance (~2 mm). The details of the transverse thermopower measurements are shown in Ref. 21.

Figure 3(c) shows the $H$ dependence of the transverse electric field $E_{\text{M}}^{y}$ normalized by $\nabla T$ in the CMG/Si hybrid bulk materials for various values of the CMG thickness $L_{\text{M}}^{z}$, measured in the open-circuit condition along the $y$ direction. All the CMG/Si hybrid bulk materials exhibit clear transverse thermopower showing the $H$-odd dependence, of which the magnitude saturates around the saturation field of CMG. The magnetization-dependent component of the transverse thermopower, $S_{\text{tot}}^{y}$, was estimated by extrapolating the linear fitting of the $H$-$E_{\text{M}}^{y}/\nabla T$ curve in the high field range (1.60 T < $\mu_0|H|$ < 1.95 T) to zero field, where $\mu_0$ is the vacuum permeability. We found that the magnitude of $S_{\text{tot}}^{y}$ monotonically increases with decreasing $L_{\text{M}}^{z}$., i.e., increasing $r$ [Figs. 3(d) and (e)], which is consistent with the feature of STTG. In the CMG/Si hybrid bulk material with largest (smallest) $r$, $S_{\text{tot}}^{y}$ was estimated to be 16.0 μV/K (6.7 μV/K), which is several times larger than (comparable to) $S_{\text{ANE}}$ of CMG [compare the red data points showing $S_{\text{tot}}^{y}$ and gray lines showing $S_{\text{ANE}}$ in Figs. 3(d) and (e)]. The $r$ dependence of $S_{\text{tot}}^{y}$ is quantitatively consistent with the values calculated by substituting the transport properties of CMG and Si (Table 1) into Eq. (1) [see the blue curve in Fig. 3(e)]. These results confirm that STTG can provide a significant contribution to the total transverse thermopower even in magnetic hybrid bulk materials with small $r$.

To experimentally demonstrate that magnetic hybrid bulk materials are beneficial to extract the output electrical power, we measured the transverse thermoelectric voltage $V_{\text{out}}$ with applying a load current $I_{\text{load}}$ to CMG along the $y$ direction. $I_{\text{load}}$ was applied in the direction to cancel the open-circuit $V_{out}$ and the output electrical power was estimated as $P_{out} = |V_{out}I_{load}|$. Figure 4 shows the $|I_{load}|$ dependence of $|V_{out}|$ and $P_{out}$ for the CMG/Si hybrid bulk material with $r = 18.0$ at the small temperature gradient of $\nabla T = 0.6$ K/mm. $|V_{out}|$ linearly decreases with increasing $|I_{load}|$, and $P_{out}$ exhibits the parabolic $|I_{load}|$ dependence, in a similar manner to the behaviors of transverse thermoelectric modules.[25] The maximum extractable electrical power of ~5.3 nW was obtained at $|I_{load}| = 0.27$ mA for the CMG/Si hybrid bulk material, while that of ~0.7 nW was obtained at 0.10 mA for the plain CMG slab with the same thickness in the absence of the electrical connection to Si. This result indicates the significant contribution of STTG to enhance the output electrical power. Importantly, the $P_{out}$ value for the CMG/Si hybrid bulk material is three orders of magnitude larger than that estimated for the thin-film-based hybrid materials used in Ref. 21 under the same temperature gradient, although $S_{tot}^y$ for the bulk-based sample is much smaller than that for the thin-film-based sample (note that the output electrical power of the CMG-film/Si-slab sample with much larger $r$ exhibits only a few pW, although the data are not shown in Ref. 21). This is due to the facts that the bulk-based sample with large $L_M^z$ has much smaller internal resistance than the thin-film-based sample [Eq. (3)], and the maximum $P_{out}$ occurs at larger $I_{load}$, confirming the usability of magnetic hybrid bulk materials for extracting electrical power through STTG. It should be noted that the $I_{load}$ application and $V_{out}$ measurements were performed with pinpoint contact probes attached to the center of the sample, as depicted in Fig. 3(a). Therefore, the electrical power was extracted only from the area around the center; if the output electrical power could be extracted from the whole of the sample, it would be much greater. In fact, the observed $P_{out}$ is an order of magnitude smaller than the electrical power expected from Eq. (4) [note that the resistance along the $y$ direction measured with the pinpoint contact probes, estimated from the slope of the $|I_{load}|$-$|V_{out}|$ curves in Fig. 4, is ~0.07 Ω but the actual sample resistance is much smaller]. Nevertheless, the data in Fig. 4 is sufficient to demonstrate that magnetic hybrid bulk materials are more useful for extracting the electrical power than thin-film-based materials. To further increase the output electrical power, it is necessary not only to find the best combination of magnetic and thermoelectric materials but also to optimize the dimensions and aspect ratio of the hybrid material, contact electrodes, and thermal boundary conditions.[15,16] Although we have focused only on the output electrical power and the Si substrate used here has high thermal conductivity, to optimize the figure of merit and efficiency of STTG devices, the thermal conductivity of hybrid materials should also be reduced by changing the material combination or by applying phonon engineering.[26]

In conclusion, we investigated STTG in the hybrid bulk materials comprising the ferromagnetic CMG slab and thermoelectric Si slab. The transverse thermopower in the CMG/Si hybrid bulk material in the open circuit condition reaches 16.0 μV/K at room temperature owing to the STTG contribution even with the small size ratio. Although this transverse thermopower is smaller than the value for the previously reported thin-film-based hybrid materials, it is still much larger than the anomalous Nernst coefficient of CMG. Our experiments demonstrate that magnetic hybrid bulk materials are suitable for extracting electrical power owing to their small internal resistance. This demonstration will invigorate materials science and device engineering studies on STTG for developing thermal energy harvesters based on transverse thermoelectrics.

The authors thank K. Suzuki and M. Isomura for technical supports. This work was supported by CREST "Creation of Innovative Core Technologies for Nano-enabled Thermal Management" (JPMJCR17I1) and ERATO "Magnetic Thermal Management Materials" (JPMJER2201) from JST, Japan, Grant-in-Aid for Scientific Research (S) (18H05246) from JSPS KAKENHI, Japan, "Mitou challenge 2050" (P14004) from NEDO, Japan, and NEC Corporation.

DATA AVAILABILITY

The data that support the findings of this study are available from the corresponding author upon reasonable request.

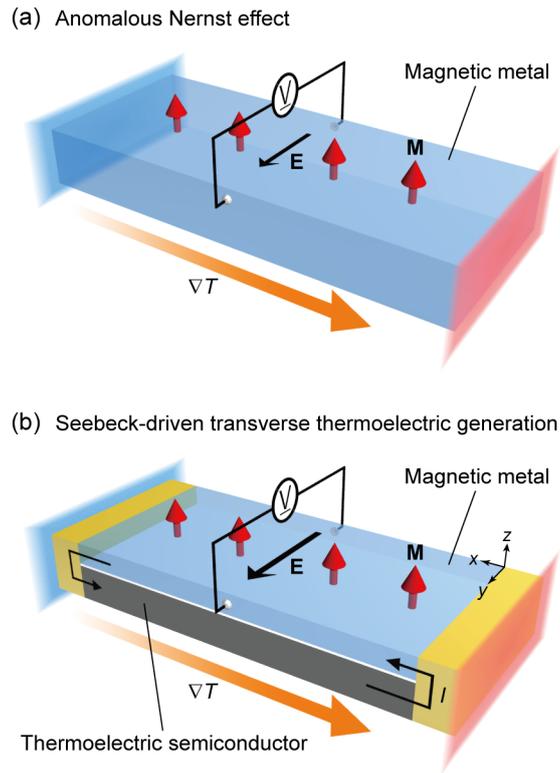

**FIG. 1.** Schematics of the anomalous Nernst effect in a magnetic metal (a) and the Seebeck-driven transverse thermoelectric generation in magnetic/thermoelectric hybrid materials (b). **E**, **M**, $\nabla T$, and *I* denote the electric field, magnetization vector, temperature gradient, and charge current in the hybrid material generated due to the difference in the Seebeck coefficient between the magnetic metal and the thermoelectric semiconductor, respectively.

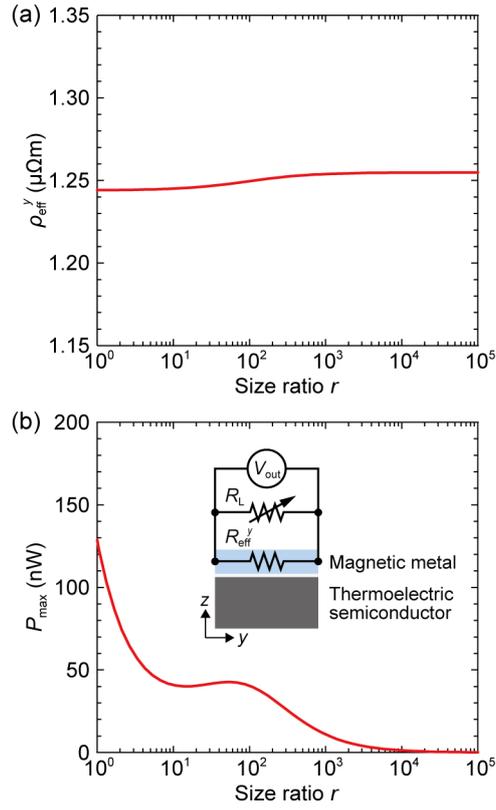

**FIG. 2.** (a) Size ratio $r$ dependence of the effective electrical resistivity along the $y$ direction $\rho_{\text{eff}}^{y}$ for the CMG/Si hybrid material, calculated from Eq. (2) with the parameters shown in Table 1. (b) $r$ dependence of the maximum extractable power $P_{\text{max}}$ for the CMG/Si hybrid material, calculated from Eq. (4) with the parameters shown in Table 1. $P_{\text{max}}$ can be estimated by measuring the output transverse thermoelectric voltage $V_{\text{out}}$ when the load resistance $R_{\text{L}}$ is matched with the effective electrical resistance along the $y$ direction $R_{\text{eff}}^{y}$ [Eq. (3)].

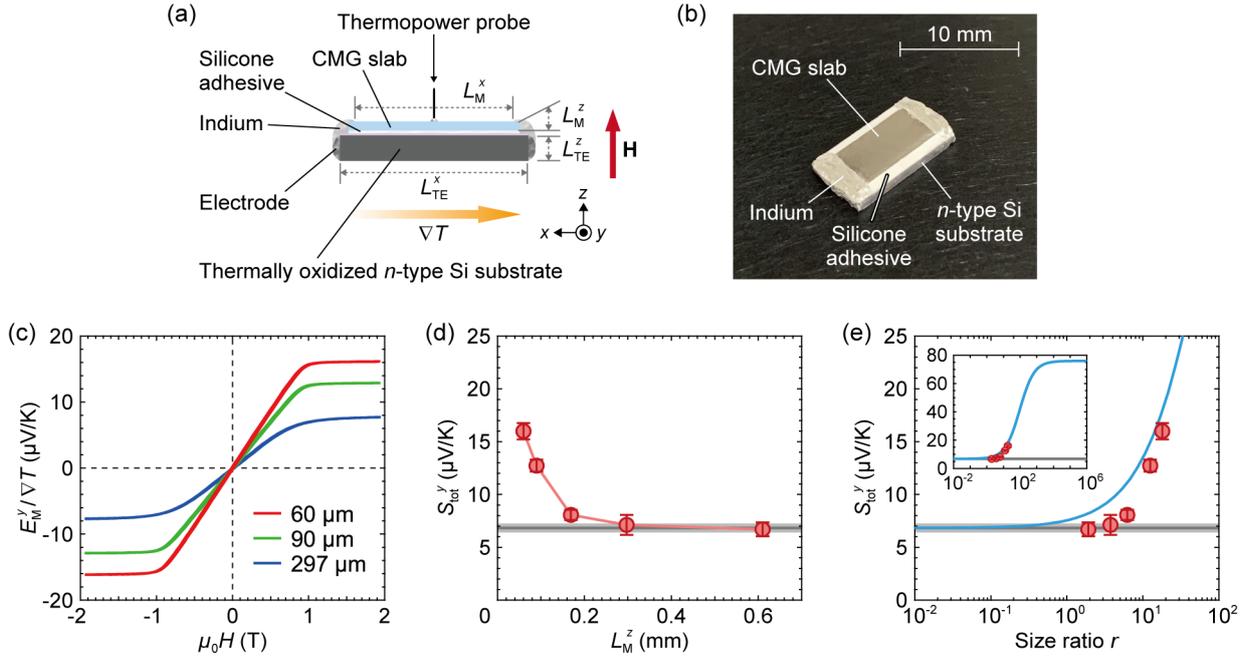

**FIG. 3.** (a) Schematic of the CMG/Si hybrid bulk material used in this study. $L^x_{\mathrm{M(TE)}}$ and $L^z_{\mathrm{M(TE)}}$ denote the lengths of the CMG (Si) slab along the $x$ and $z$ directions, respectively. $L^x_{\mathrm{M}}$ is determined as the distance between the two indium contacts. **H** denotes the direction of the external magnetic field. (b) Photograph of the CMG/Si hybrid bulk material. (c) Magnetic field $H$ dependence of the transverse electric field in the CMG slab $E^y_{\mathrm{M}}$ normalized by $\nabla T$ for various values of $L^z_{\mathrm{M}}$. $\mu_0$ denotes the vacuum permeability. (d) $L^z_{\mathrm{M}}$ dependence of the transverse thermopower $S^y_{\mathrm{tot}}$ in the CMG/Si hybrid bulk materials. (e) $r$ dependence of $S^y_{\mathrm{tot}}$. The red data points show the experimentally observed $S^y_{\mathrm{tot}}$ values for the CMG/Si hybrid bulk materials. The blue curve shows the transverse thermopower calculated using Eq. (1). The inset to (e) shows the same data for the wider $r$ and $S^y_{\mathrm{tot}}$ ranges. The gray lines in (d) and (e) show the anomalous Nernst coefficient of the plain CMG slab.

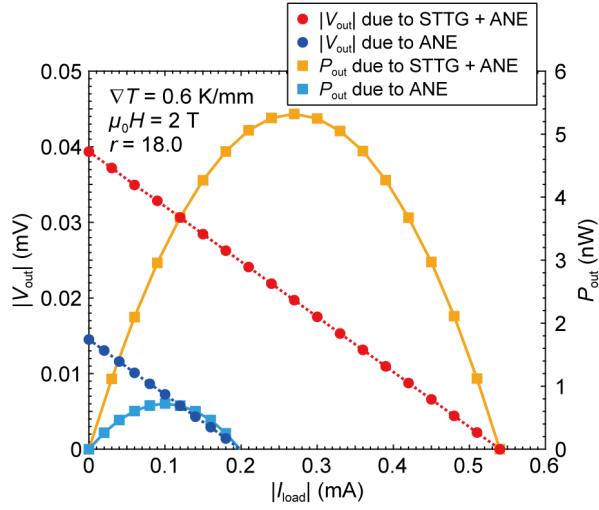

**FIG. 4.** Load current magnitude $|I_{\text{load}}|$ dependence of the output voltage magnitude $|V_{\text{out}}|$ and electrical power $P_{\text{out}}$ along the $y$ direction for the CMG/Si hybrid bulk material with $r = 18.0$, used in Fig. 3, at $\nabla T = 0.6$ K/mm and $\mu_0 H = 2$ T. The magnetization of CMG aligns along the $z$ direction because the applied external magnetic field is much greater than the saturation field of CMG.

**Table 1.** Transport properties of CMG and $n$-type Si used in this study. $\rho_{\text{M(TE)}}$ and $S_{\text{M(TE)}}$ denote the longitudinal electrical resistivity and Seebeck coefficient of CMG (Si), respectively, which were simultaneously measured with the Seebeck coefficient/electric resistance measurement system (ZEM-3, ADVANCE RIKO, Inc). $\rho_{\text{AHE}}$ and $S_{\text{ANE}}$ denote the anomalous Hall resistivity and anomalous Nernst coefficient of CMG, respectively.

|  | CMG | $n$-type Si |
|---|---|---|
| $\rho_{\text{M}}$ or $\rho_{\text{TE}}$ ($\Omega$m) | $1.24 \times 10^{-6}$ | $1.17 \times 10^{-4}$ |
| $S_{\text{M}}$ or $S_{\text{TE}}$ (V/K) | $-33.2 \times 10^{-6}$ | $-7.75 \times 10^{-4}$ |
| $\rho_{\text{AHE}}$ ($\Omega$m) | $0.12 \times 10^{-6}$ | - |
| $S_{\text{ANE}}$ (V/K) | $6.8 \times 10^{-6}$ | - |